\documentclass[12pt]{article}
\usepackage{epsf}
\usepackage{amsmath}
\def\hybrid{\topmargin 0pt      \oddsidemargin 0pt
	\headheight 0pt \headsep 0pt
	\textheight 9in         
	\textwidth 6.25in       
	\marginparwidth .875in
	\parskip 5pt plus 1pt   \jot = 1.5ex}

\catcode`\@=11
\def\marginnote#1{}
\newcount\hour
\newcount\minute
\newtoks\amorpm
\hour=\time\divide\hour by60
\minute=\time{\multiply\hour by60 \global\advance\minute by-\hour}
\edef\standardtime{{\ifnum\hour<12 \global\amorpm={am}%
	\else\global\amorpm={pm}\advance\hour by-12 \fi
	\ifnum\hour=0 \hour=12 \fi
	\number\hour:\ifnum\minute<10 0\fi\number\minute\the\amorpm}}
\edef\militarytime{\number\hour:\ifnum\minute<10 0\fi\number\minute}

\def\draftlabel#1{{\@bsphack\if@filesw {\let\thepage\relax
   \xdef\@gtempa{\write\@auxout{\string
      \newlabel{#1}{{\@currentlabel}{\thepage}}}}}\@gtempa
   \if@nobreak \ifvmode\nobreak\fi\fi\fi\@esphack}
	\gdef\@eqnlabel{#1}}
\def\@eqnlabel{}
\def\@vacuum{}
\def\draftmarginnote#1{\marginpar{\raggedright\scriptsize\tt#1}}

\def\draft{\oddsidemargin -.5truein
	\def\@oddfoot{\sl preliminary draft \hfil
	\rm\thepage\hfil\sl\today\quad\militarytime}
	\let\@evenfoot\@oddfoot \overfullrule 3pt
	\let\label=\draftlabel
\let\marginnote=\draftmarginnote
	\let\marginnote=\draftmarginnote
   \def\@eqnnum{(\theequation)\rlap{\kern\marginparsep\tt\@eqnlabel}%
\global\let\@eqnlabel\@vacuum}  }

\def\numberbysection{\@addtoreset{equation}{section}
\def\theequation{\thesection.\arabic{equation}}}

\def\underline#1{\relax\ifmmode\@@underline#1\else
	$\@@underline{\hbox{#1}}$\relax\fi}

\def\titlepage{\@restonecolfalse\if@twocolumn\@restonecoltrue\onecolumn
     \else \newpage \fi \thispagestyle{empty}\c@page\z@
	\def\thefootnote{\fnsymbol{footnote}} }

\def\endtitlepage{\if@restonecol\twocolumn \else  \fi
	\def\thefootnote{\arabic{footnote}}
	\setcounter{footnote}{0}}  
\catcode`@=12
\relax

\def\beq{\begin{equation}}
\def\eeq{\end{equation}}
\def\bea{\begin{eqnarray}}
\def\eea{\end{eqnarray}}
\def\nn{\nonumber}

\relax
\hybrid

\begin{document}

\begin{titlepage}
\begin{center}
May~2012 \hfill . \\[.5in]
{\large\bf  Two more solutions for the parafermionic chiral algebra $Z_{3}$ with the dimension of the principal parafermionic fields, $\psi(z)$, $\psi^{+}(z)$, $\Delta_{\psi}=8/3$.}
\\[.5in] 
{\bf Vladimir S.~Dotsenko}\\[.2in]
{\it LPTHE, CNRS, Universit{\'e} Pierre et Marie Curie, Paris VI, UMR 7589\\
               4 place Jussieu,75252 Paris Cedex 05, France.}\\[.2in]
               
 \end{center}
 
\underline{Abstract.}

In this paper, which is the second one in a series of two papers, we shall present two more solutions, non-minimal ones, for the $Z_{3}$ parafermionic chiral algebra with $\Delta_{\psi}=\Delta_{\psi^{+}}=8/3$, $\psi(z)$, $\psi^{+}(z)$ being the principal parafermionic fields.

\end{titlepage}

\newpage

\numberwithin{equation}{section}

\section{Introduction of the chiral algebra.}

Our methods of searching for associative chiral algebras, of parafermionic type, have been presented in much detail in the previous paper [1]. For this reason the presentation in the present paper will be much less detailed.

The parafermionic algebras, which we are going to describe in this paper, are closed by the following (chiral) fields:
\beq
\psi(z), \,\, \psi^{+}(z), \,\, \tilde{\psi}(z), \,\, \tilde{\psi}^{+}(z), \,\, U(z), \,\,
 B(z), \,\, W(z)\label{eq1.1}
\eeq
having the dimensions :
\bea
\Delta_{\psi}=\Delta_{\psi^{+}}=\frac{8}{3},\quad \Delta_{\tilde{\psi}}=\Delta_{\tilde{\psi}^{+}}=\Delta_{\psi}+2=\frac{14}{3},\nn\\
\Delta_{U}=3,\quad \Delta_{B}=4,\quad\Delta_{W}=5\label{eq1.2}
\eea
Their operator product expansions (OPE) are of the form :
\beq
\psi\psi=\lambda[\psi^{+}]+\zeta[\tilde{\psi}^{+}]\label{eq1.3}
\eeq
\beq
\psi\psi^{+}=[I]+\alpha[U]+\gamma[B]+\delta[W]\label{eq1.4}
\eeq
\beq
\psi\tilde{\psi}=\zeta[\psi^{+}]+\eta[\tilde{\psi}^{+}]\label{eq1.5}
\eeq
\beq
\psi^{+}\tilde{\psi}=\kappa[U]+\mu[B]+\nu[W]\label{eq1.6}
\eeq
\beq
\tilde{\psi}\tilde{\psi}=\eta[\psi^{+}]+\tilde{\lambda}[\tilde{\psi}^{+}]\label{eq1.7}
\eeq
\beq
\tilde{\psi}\tilde{\psi^{+}}=[I]+\tilde{\alpha}[U]+\tilde{\gamma}[B]+\tilde{\delta}[W]\label{eq1.8}
\eeq
\beq
\psi U=-\alpha[\psi]+\kappa[\tilde{\psi}]\label{eq1.9}
\eeq
\beq
\psi B=\gamma[\psi]+\mu[\tilde{\psi}]\label{eq1.10}
\eeq
\beq
\psi W=-\delta[\psi]+\nu[\tilde{\psi}]\label{eq1.11}
\eeq
\beq
\tilde{\psi}U=\kappa[\psi]-\tilde{\alpha}[\tilde{\psi}]\label{eq1.12}
\eeq
\beq
\tilde{\psi}B=\mu[\psi]+\tilde{\gamma}[\tilde{\psi}]\label{eq1.13}
\eeq
\beq
\tilde{\psi}W=\nu[\psi]-\tilde{\delta}[\tilde{\psi}]\label{eq1.14}
\eeq
\beq
UU=[I]+a[B]\label{eq1.15}
\eeq
\beq
UB=a[U]+e[W]\label{eq1.16}
\eeq
\beq
UW=e[B]\label{eq1.17}
\eeq
\beq
BB=[I]+b[B]\label{eq1.18}
\eeq
\beq
BW=e[U]+d[W]\label{eq1.19}
\eeq
\beq
WW=[I]+d[B]\label{eq1.20}
\eeq
In the equations above, $[\psi^{+}]$ represents the Virasoro algebra series
of $\psi^{+}$, and similarly for other operators in the r.h.s. of (\ref{eq1.3})-(\ref{eq1.20}). We are using the same notations as in [1].

In order to give examples, the compact forms of the OPEs in (\ref{eq1.3})-(\ref{eq1.6}) should be viewed, explicitly, as follows :
\bea
\psi(z')\psi(z)=\frac{1}{(z'-z)^{\Delta_{\psi}}}\{\lambda\psi^{+}(z)+(z'-z)\beta^{(1)}_{\psi\psi,\psi^{+}}\cdot\partial\psi^{+}(z)\nn\\
+(z'-z)^{2}(\lambda\beta^{(11)}_{\psi\psi,\psi^{+}}\cdot\partial^{2}\psi^{+}(z)+\lambda\beta^{(2)}_{\psi\psi,\psi^{+}}\cdot L_{-2}\psi^{+}(z)+\zeta\tilde{\psi}^{+}(z))+...\}\label{eq1.21}
\eea
\bea
\psi(z')\psi^{+}(z)=\frac{1}{(z'-z)^{2\Delta_{\psi}}}\{1+(z'-z)^{2}\frac{2\Delta_{\psi}}{c}T(z)\nn\\
+(z'-z)^{3}\cdot(\frac{\Delta_{\psi}}{c}\partial T(z)+\alpha\cdot U(z))\nn\\
+(z'-z)^{4}\cdot(\beta^{(112)}_{\psi\psi^{+},I}\cdot\partial^{2}T(z)+\beta^{(22)}_{\psi\psi^{+},I}\cdot\wedge(z)+\alpha\beta^{(1)}_{\psi\psi^{+},U}\cdot\partial U(z)+\gamma\cdot B(z))\nn\\
+(z'-z)^{5}(\beta^{(1112)}_{\psi\psi^{+},I}\cdot\partial^{3}T(z)+\beta^{(122)}_{\psi\psi^{+},I}\partial\wedge(z)\nn\\
+\alpha\beta^{(11)}_{\psi\psi^{+},U}\partial^{2}U(z)+\alpha\beta^{(2)}_{\psi\psi^{+},U}\cdot L_{-2}U(z)\nn\\
+\gamma\beta^{(1)}_{\psi\psi^{+},B}\cdot\partial B(z)+\delta\cdot W(z))+...\} \label{eq1.22}
\eea
\bea
\psi(z')\tilde{\psi}(z)=\frac{1}{(z'-z)^{\Delta_{\tilde{\psi}}}}\{\zeta\psi^{+}(z)+(z'-z)\zeta\beta^{(1)}_{\psi\tilde{\psi},\psi^{+}}\cdot\partial\psi^{+}(z)\nn\\
+(z'-z)^{2}(\zeta\beta^{(11)}_{\psi\tilde{\psi},\psi^{+}}\partial^{2}\psi^{+}(z)+\zeta\beta^{(2)}_{\psi\tilde{\psi},\psi^{+}}L_{-2}\psi^{+}(z)+\eta\tilde{\psi}^{+}(z))+...\}\label{eq1.23}
\eea
\bea
\psi^{+}(z')\tilde{\psi}(z)=\frac{1}{(z'-z)^{2\Delta_{\psi}-1}}\{\kappa U(z)+(z'-z)(\kappa\beta^{(1)}_{\psi^{+}\tilde{\psi},U}\cdot\partial U(z)+\mu B(z))\nn\\
+(z'-z)^{2}(\kappa\beta^{(11)}_{\psi^{+}\tilde{\psi},U}\cdot\partial^{2}U(z)+\kappa\beta^{(2)}_{\psi^{+}\tilde{\psi},U}L_{-2}U(z)\nn\\
+\mu\beta^{(1)}_{\psi^{+}\tilde{\psi},B}\partial B(z)+\nu W(z))+...\}\label{eq1.24}
\eea

As has been explained in [1], for the algebra with the conformal dimensions
of the principal parafermionic fields $\Delta_{\psi}=8/3$, we have to make explicit the expansions in the $Z_3$ neutral sector up to level 5,
and up to level 2 in the $Z_3$ charged sector, 
cf. (\ref{eq1.21})-(\ref{eq1.24}).

Alternatively, the operator algebra constants in (\ref{eq1.3})-(\ref{eq1.20}) can be defined by the 3-point functions as follows :
\bea
<\psi\psi\psi>=\lambda, \,\,
<\psi\psi\tilde{\psi}>=\zeta, \,\,
<\psi\tilde{\psi}\tilde{\psi}>= \eta, \,\,
<\tilde{\psi}\tilde{\psi}\tilde{\psi}>=\tilde{\lambda},\nn\\
<\psi\psi^{+}U>= \alpha, \,\,
<\psi\psi^{+} B>=\gamma, \,\,
<\psi\psi^{+}W>=\delta, \nn\\
<\psi^{+}\tilde{\psi}U>=\kappa, \,\,
<\psi^{+}\tilde{\psi}B>=\mu, \,\,
<\psi^{+}\tilde{\psi}W>= \nu, \nn\\
<\tilde{\psi}\tilde{\psi}^{+}U>= \tilde{\alpha}, \,\,
<\tilde{\psi}\tilde{\psi}^{+}B>= \tilde{\gamma}, \,\,
<\tilde{\psi}\tilde{\psi}^{+}W>=\tilde{\delta}, \nn\\
<BUU>= a, \,\,
<BBB>=b, \,\,
<WWB>= d, \,\,
<WBU>= e\label{eq1.41}
\eea
Here $<\psi\psi\psi>=<\psi(\infty)\psi(1)\psi(0)>$, etc. .

We observe that the bosonic operators $U$ and $W$, with the odd conformal dimensions, have to be odd with respect to the $Z_{3}$ conjugation :
\beq
U^{+}=-U,\quad W^{+}=-W\label{eq1.42}
\eeq
while B, with dimension 4, have to be even :
\beq
B^{+}=B\label{eq1.43}
\eeq
This follows from the OPE (\ref{eq1.22}) : if we continue analytically
$\psi^{+}(z)$ around $\psi(z')$, $\psi(z')\psi^{+}(z)\rightarrow\psi^{+}(z)\psi(z')$, the l.h.s. takes the form of the initial product $Z_{3}$ conjugated, while in the r.h.s. the terms $\sim U$, $W$ will change their signs (the phase factors in front, in the l.h.s. and the r.h.s., are assumed to compensate each other [1]).

E.g. this implies that, from $<\psi\psi^{+}U>= \alpha$, eq.(\ref{eq1.41}), by the $Z_{3}$ conjugation, one obtains
\beq
<\psi^{+}\psi U>=-\alpha\label{eq1.44}
\eeq
while, from $<\psi\psi^{+} B>=\gamma$, one finds that
\beq
<\psi^{+}\psi B>=\gamma\label{eq1.45}
\eeq

This implies also that the signs will change, e.g. in (\ref{eq1.22}), of the terms $\sim\alpha$, $\delta$, if we develop the product $\psi^{+}(z')\psi(z)$ instead of $\psi(z')\psi^{+}(z)$.

Similarly, the signs of terms $\sim\kappa$, $\nu$ in (\ref{eq1.24}) will change under the $Z_{3}$ conjugation, giving the decomposition of the product $\psi(z')\tilde{\psi}^{+}(z)$.

The odd nature of $U$ and $W$ implies also that $<UUU>=0$, $<WWW>=0$, $<BBU>=0$, and so on, thus reducing considerably the number of purely bosonic constants.

As has already been stated in [1], for the chiral algebra (\ref{eq1.3})-(\ref{eq1.20}), 
of the chiral fields in (\ref{eq1.1}), we have found two different solutions 
of the associativity constraints : two different sets of values for the constants (\ref{eq1.41}), with the central charge staying unconstrained, remain  
a free parameter. These solutions will be presented in the next Section,
together with some indications to the methods that we have used for their derivations.

\section{Determination of the operator algebra constants.}

The method, which uses the analysis of triple products, has been described in detail in [1]. Like in [1] we start with the "light" triple products 
and we proceed
to the "heavier" ones, heavier in terms of the values of dimensions of the fields in the product. Saying it differently, we start with the principal fields, 
$\psi$, $\psi^{+}$ (which generate the whole algebra) and proceed by including the secondary fields, 
$\tilde{\psi}$, $\tilde{\psi}^{+}$, $U$, $B$, $W$. Eventually, the products which includes too many of secondary fields (become too heavy), they become irrelevant, in the sense that they do not produce equations on operator algebra constants. So that there is a limited number of triple products which have to analyzed.

We proceed like in [1] and we get the triple products, that follow, 
which produce equations en the constants (\ref{eq1.41}).
\beq
\mbox{\underline{1. $\psi\psi\psi^{+}\rightarrow\psi$;}}\quad
 <\psi^{+}\psi\psi\psi^{+}>\label{eq2.1}
\eeq
3 equations,
\beq
1.1:\quad \alpha^{2},\zeta^{2},\mbox{\underline{$\lambda^{2}$}}\label{eq2.2}
\eeq
\beq
1.2:\quad \alpha^{2},\zeta^{2},\lambda^{2},
\mbox{\underline{$\gamma^{2}$}}\label{eq2.3}
\eeq
\beq
1.3:\quad \alpha^{2},\zeta^{2},\lambda^{2},\gamma^{2},
\mbox{\underline{$\delta^{2}$}}\label{eq2.4}
\eeq
Above, in (\ref{eq2.1}), we indicate the channel that we have looked at, in the expansion of the triple product, and the corresponding to that channel 
4-point function.

The resulting equations are listed in the Appendix A and 
in the Appendix B we give an example of their derivation.

In (\ref{eq2.2}-(\ref{eq2.4}) we give our numeration of the equations (of the Appendix A); we also indicate the constants which enter the corresponding equations. This will allow us to describe the procedure which we have followed to determine all the constants. For that purpose, we also underline particular constants in (\ref{eq2.2})-(\ref{eq2.4}). Those are the constants which will be determined by the corresponding equation. We shall make it more precise further down.
\beq
\mbox{\underline{2. $\psi\psi\tilde{\psi}^{+}\rightarrow\partial^{2}\psi,  L_{-2}\psi,  \tilde{\psi}$;}}\quad
<\psi^{+}\psi\psi\tilde{\psi}^{+}>, <\tilde{\psi}^{+}\psi\psi\tilde{\psi}^{+}>\label{eq2.5}
\eeq
3 equations,
\beq
2.1:\quad \lambda,\zeta,\mbox{\underline{$\eta$}},\alpha,\gamma,\delta,\kappa,\mu,\nu\label{eq2.6}
\eeq
\beq
2.2:\quad \lambda,\zeta,\eta,\alpha,\gamma,\delta,\kappa,\mu,\nu\label{eq2.7}
\eeq
\beq
2.3:\quad \lambda,\zeta,\eta,\mbox{\underline{$\tilde{\lambda}$}},
\kappa^{2},\mu^{2},\nu^{2}\label{eq2.8}
\eeq
\beq
\mbox{\underline{3. $\psi\psi\psi\rightarrow\partial^{3}T, \partial\wedge, \partial^{2}U, L_{-2}U, \partial B, W$;}}\quad
<U\psi\psi\psi>, <B\psi\psi\psi>, <W\psi\psi\psi>\label{eq2.9}
\eeq
3 equations,
\beq
3.1:\quad\zeta,\lambda,\alpha,\mbox{\underline{$\kappa$}}\label{eq2.10}
\eeq
\beq
3.2:\quad\zeta,\lambda,\gamma,\mbox{\underline{$\mu$}}\label{eq2.11}
\eeq
\beq
3.3:\quad\zeta,\lambda,\delta,\mbox{\underline{$\nu$}}\label{eq2.12}
\eeq
\beq
\mbox{\underline{4. $\psi\psi^{+}U\rightarrow\partial^{3}T, \partial\wedge, \partial^{2}U, L_{-2}U, \partial B, W$;}}
<U\psi\psi^{+}U>, <B\psi\psi^{+}U>, <W\psi\psi^{+}U>\label{eq2.13}
\eeq
5 equations,
\beq
4.1:\quad\alpha,\delta,\gamma,a,\mbox{\underline{$e$}}\label{eq2.14}
\eeq
\beq
4.2:\quad\alpha^{2},\gamma,\kappa^{2},\mbox{\underline{$a$}}\label{eq2.15}
\eeq
\beq
\mbox{\underline{$4.3$}}:\quad\alpha^{2},\gamma,\kappa^{2},a\label{eq2.16}
\eeq
\beq
4.4.:\quad\alpha,\gamma,\delta,\kappa,\mu,a,e\label{eq2.17}
\eeq
\beq
\mbox{\underline{$4.5$}}:\quad\alpha,\gamma,\delta,\kappa,\nu,e\label{eq2.18}
\eeq
\beq
\mbox{\underline{5. $\psi\psi^{+}B\rightarrow\partial^{3}T, \partial\wedge, \partial^{2}U, L_{-2}U, \partial B, W$;}}
<U\psi\psi^{+}B>,<B\psi\psi^{+}B>,<W\psi\psi^{+}B>\label{eq2.19}
\eeq
4 equations,
\beq
5.1:\quad\alpha,\gamma,\delta,\kappa,\mu,a,e\label{eq2.20}
\eeq
\beq
5.2:\quad\alpha,\gamma,\delta,\kappa,\mu,a,e\label{eq2.21}
\eeq
\beq
5.3:\quad\gamma,\mu,\mbox{\underline{$b$}}\label{eq2.22}
\eeq
\beq
5.4.:\quad\alpha,\gamma,\delta,\mu,\nu,\mbox{\underline{$d$}},e\label{eq2.23}
\eeq
\beq
\mbox{\underline{6. $\psi\psi^{+}W\rightarrow\partial^{3}T, \partial\wedge, \partial^{2}U, L_{-2}U, \partial B, W$;}}
<U\psi\psi^{+}W>, <B\psi\psi^{+}W>, <W\psi\psi^{+}W>\label{eq2.24}
\eeq
1 equation is produced, corresponding to the projection $\psi\psi^{+}W
\rightarrow\partial B$; this equation is identical to the eq.5.4 above.
\beq
\mbox{\underline{7. $\psi^{+}\tilde{\psi}U\rightarrow\partial^{3}T, \partial\wedge, \partial^{2}U, L_{-2}U,\partial B, W$;}}
<U\psi^{+}\tilde{\psi}U>, <B\psi^{+}\tilde{\psi}U>, <W\psi^{+}\tilde{\psi}U>\label{eq2.25}
\eeq
3 equations,
\beq
7.1:\quad\alpha,\kappa,\mu,a,\mbox{\underline{$\tilde{\alpha}$}}\label{eq2.26}
\eeq
\beq
7.2:\quad\alpha,\gamma,\kappa,\mu,\nu,a,e,\tilde{\alpha},
\mbox{\underline{$\tilde{\gamma}$}}\label{eq2.27}
\eeq
\beq
7.3:\quad\alpha,\delta,\kappa,\mu,\nu,e,\tilde{\alpha},
\mbox{\underline{$\tilde{\delta}$}}\label{eq2.28}
\eeq
\beq
\mbox{\underline{8.  $\psi\psi\tilde{\psi}\rightarrow\partial^{3}T, \partial\wedge, \partial^{2}U, L_{-2}U, \partial B, W$;}}
<U\psi\psi\tilde{\psi}>, <B\psi\psi\tilde{\psi}>, <W\psi\psi\tilde{\psi}>\label{eq2.29}
\eeq
4 equations,
\beq
8.1:\quad\alpha,\zeta,\lambda,\eta,\kappa,\tilde{\alpha}\label{eq2.30}
\eeq
\beq
8.2:\quad\alpha,\zeta,\lambda,\eta,\kappa,\tilde{\alpha}\label{eq2.31}
\eeq
\beq
8.3:\quad\zeta,\lambda,\gamma,\eta,\mu,\tilde{\gamma}\label{eq2.32}
\eeq
\beq
8.4.:\quad\zeta,\delta,\lambda,\eta,\nu,\tilde{\delta}\label{eq2.33}
\eeq
\beq
\mbox{\underline{9. $\psi\psi^{+}\tilde{\psi}\rightarrow\partial^{2}\psi, L_{-2}\psi, \tilde{\psi}$;}}\quad
<\psi^{+}\psi\psi^{+}\tilde{\psi}>, <\tilde{\psi}^{+}\psi\psi^{+}\tilde{\psi}>\label{eq2.34}
\eeq
5 equations,
\beq
9.1:\quad\alpha,\zeta,\delta,\lambda,\gamma,\kappa,\mu,\nu\label{eq2.35}
\eeq
\beq
9.2:\quad\alpha,\zeta,\delta,\lambda,\eta,\gamma,\kappa,\mu,\nu\label{eq2.36}
\eeq
\beq
9.3:\quad\alpha,\zeta,\delta,\lambda,\eta,\gamma,\kappa,\mu,\nu\label{eq2.37}
\eeq
\beq
9.4.:\quad\alpha,\zeta,\delta,\gamma,\kappa,\mu,\nu,
\tilde{\alpha},\tilde{\gamma},\tilde{\delta}\label{eq2.38}
\eeq
\beq
9.5:\quad\alpha,\zeta,\delta,\gamma,\eta,\kappa,\mu,\tilde{\alpha},\tilde{\gamma},\tilde{\delta}\label{eq2.39}
\eeq

\vskip1cm

Still there remain the following triple products which are relevant:
\beq
\mbox{\underline{10. $\psi \psi U\rightarrow\partial^{2}\psi^{+}, L_{-2}\psi^{+},\tilde{\psi}^{+}$;}}
<\psi\psi\psi U>,<\tilde{\psi}\psi\psi U>\label{eq2.40}
\eeq
\beq
\mbox{\underline{11. $\psi\psi B\rightarrow\partial^{2}\psi^{+}, L_{-2}\psi^{+}, \tilde{\psi}^{+}$;}}
<\psi\psi\psi B>,<\tilde{\psi}\psi\psi B>\label{eq2.41}
\eeq
\beq
\mbox{\underline{12. $\psi\psi W\rightarrow\partial^{2}\psi^{+},L_{-2}\psi^{+},\tilde{\psi}^{+}$;}}
<\psi\psi\psi W>,<\tilde{\psi}\psi\psi W>\label{eq2.42}
\eeq
By looking at the corresponding 4-point functions, one could see that, with respect to equations that they will produce, (\ref{eq2.40})-(\ref{eq2.42}) are equivalent (\ref{eq2.9}) and (\ref{eq2.29}).
\beq
\mbox{\underline{13. $\psi\tilde{\psi}U\rightarrow\partial^{2}\psi^{+},L_{-2}\psi^{+},\tilde{\psi}^{+}$;}}<\psi\psi\tilde{\psi}U>,<\tilde{\psi}\psi\psi U>\label{eq2.43}
\eeq
3 equations,
\beq
13.1:\quad \alpha,\zeta,\lambda,\eta,\kappa,\tilde{\alpha}\label{eq2.43.1}
\eeq
\beq
13.2:\quad \alpha,\zeta,\lambda,\eta,\kappa,\tilde{\alpha}\label{eq2.43.2}
\eeq
\beq
13.3:\quad \alpha,\zeta,\eta,\kappa,\tilde{\lambda},\tilde{\alpha}\label{eq2.43.3}
\eeq
This product will be analyzed  later, after the calculation of the constants. We shall examine it, in some detail, in the Appendix B, 
as an example for deriving equations.
\beq
\mbox{\underline{14. $\psi UU\rightarrow\partial^{2}\psi,L_{-2}\psi,\tilde{\psi}$;}}<\psi^{+}\psi UU>,<\tilde{\psi}^{+}\psi UU>\label{eq2.43.4}
\eeq
By the corresponding 4-point functions, the equations produced by this product should be equivalent to certain equations produced by the products 4 and 7.
\beq
\mbox{\underline{15. $UUU\rightarrow U,W$;}}<UUUU>,<WUUU>\label{eq2.43.5}
\eeq
The equations produced by this product are verified in a trivial way, for all values of the constants $a$ and $e$ which appear for this product. 
The corresponding 4 point correlation functions must be too simple 
in this case.
\beq
\mbox{\underline{16. $\psi UB\rightarrow\partial^{2}\psi,L_{-2}\psi,\tilde{\psi}$;}}<\psi^{+}\psi UB>,<\tilde{\psi}^{+}\psi UB>\label{eq2.43.6}
\eeq
Equivalent, by their 4-point functions, to the products 4 and 7 (to their particular equations).

\beq
\mbox{\underline{17. $\psi UW\rightarrow\partial^{2}\psi,L_{2}\psi,\tilde{\psi}$;}}<\psi^{+}\psi UW>,<\tilde{\psi}^{+}\psi UW>\label{eq2.43.7}
\eeq
The same as for the product 16 above.

\beq
\mbox{\underline{18. $UUB\rightarrow B$;}}<BUUB>\label{eq2.43.8}
\eeq
1 equation,
\beq
18.1:\quad a,b,e\label{eq2.43.9}
\eeq

\beq
\mbox{\underline{19. $UU\tilde{\psi}\rightarrow\partial^{2}\psi,L_{-2}\psi,\tilde{\psi}$;}}<\psi^{+}UU\tilde{\psi}>,<\tilde{\psi}^{+}UU\tilde{\psi}>\label{eq2.43.10}
\eeq

\beq
\mbox{\underline{20. $UUW\rightarrow U,W$;}}<UUUW>,<WUUW>\label{eq2.43.11}
\eeq
The products 19, 20 do not produce equations. Products 18, 19, 20 are at the edge between the products of type 1, which produce equations, and those of type 2 which do not produce equations on the constants ("being too heavy"), but for which all the matrix elements, needed to calculate the corresponding 4-point functions, get determined by the commutation relations, cf.[1].

It happens that the product 18 is on one side of the edge and the products 19, 20 find themselves on the other side.

\vskip1.5cm

\underline{To determine the values of the constants we have proceeded as follows}.

First we have defined $\lambda$ from (\ref{eq2.2}), as a function of $\alpha$ and $\zeta$, $\lambda(\alpha,\zeta)$. We have substituted it into (\ref{eq2.3}) and we have defined $\gamma$, as a function of $\alpha$ and $\zeta$, $\gamma(\alpha,\zeta)$. Next, putting $\lambda(\alpha,\zeta)$ and $\gamma(\alpha,\zeta)$ into (\ref{eq2.4}) we have defined $\delta(\alpha,\zeta)$.

Proceeding in a similar way, we have defined $\kappa(\alpha,\zeta)$, $\mu(\alpha,\zeta)$, $\nu(\alpha,\zeta)$ from the equations 
(\ref{eq2.10})-(\ref{eq2.12}). We remind that all these equations are available in the Appendix A.

Further down, from (\ref{eq2.6}) we have determined $\eta(\alpha,\zeta)$. With this value of $\eta$, and the values of the other constants, eq,(\ref{eq2.7}) gets satisfied, while eq.(\ref{eq2.8}) allows to define 
$\tilde{\lambda}(\alpha,\zeta)$.

Next, (\ref{eq2.15}) defines $a(\alpha,\zeta)$, (\ref{eq2.14}) then defines $e(\alpha,\zeta)$, and equations (\ref{eq2.22}), (\ref{eq2.23}) define $b(\alpha,b)$ and $d(\alpha,\zeta)$.

Finally, with the equations (\ref{eq2.30}), (\ref{eq2.32}), (\ref{eq2.33}) we have defined $\tilde{\alpha}(\alpha,\zeta)$, $\tilde{\gamma}(\alpha,\zeta)$, $\tilde{\delta}(\alpha,\zeta)$.

In this way we have determined 15 constants, out of a total 17, as functions of $\alpha$ and $\zeta$. It remains then to define $\alpha$ and $\zeta$, as functions of $c$, the central charge. There are many ways to do it. In the set of equations that we have obtained (listed above and in the Appendix A) there are still many that we haven't used. Finally, we did the calculations as follows.

We have substituted $\gamma(\alpha,\zeta)$, $\kappa(\alpha,\zeta)$, $a(\alpha,\zeta)$ into the equation (\ref{eq2.16}) (equation number 4.3, in the Appendix A). This gave us an equation en $\alpha$, $\zeta$. Solving it we have defined $\zeta$ as a function of $\alpha$, $\zeta(\alpha)$.

Next we have chosen eq.(\ref{eq2.18}) (equation number 4.5), to substitute all the constants, which are functions of $\alpha,\zeta$, and substitute also $\zeta(\alpha)$, just defined. This gave us an equation en $\alpha$. Solving it, we have defined $\alpha$. 

One gets actually four values of $\alpha$ which are solutions of the last equation.

As has been said above, there many ways to proceed with the above calculation. This is because the number of equations is bigger than the number of constants.

The final results have to be the same, if the solution exist, but the intermediate expressions might be very different in their complexity. For example, the equation (\ref{eq2.16}) has been chosen, after many attempts, because it gave the simplest expression of $\zeta(\alpha)$. Similarly for the choice of the equation (\ref{eq2.18}), to define $\alpha$.

The intermediate expressions are rather long, at least some of them, when the constants are defined first as functions of $\alpha$, $\zeta$. It is not reasonable to provide all of them. The final expressions are much simpler. For this reason we shall give just one intermediate formula, that for 
$\zeta(\alpha)$, solution of the equation (\ref{eq2.16}). To reproduce all the preceding formulas, for constants as functions of $\alpha,\zeta$, is straightforward in fact, having the equations and the indications that we 
have given above.

We observe also that all our calculations have been done with Mathematica.

We have obtained the following formula for $\zeta(\alpha)$ :
\beq
\zeta(\alpha)= \frac{\sqrt{-\alpha^2 (272 + 
   11 c)^2 (-2688 + (-32 + 81 \alpha^2) c)}}{\sqrt{2} \sqrt{(784 + 
    57 c) (7424 + c (-64 + 81 \alpha^2 (46 + c)))}} \label{eq2.44}
\eeq
After that, we have found the following four allowed values of $\alpha$ :
\beq
\alpha_{1}= \frac{8}{9} \sqrt{10} \sqrt{-\frac{56 + c}{c (72 + c)}}
\label{eq2.45}
\eeq
\beq
\alpha_{2}= \frac{4}{9} \sqrt{2} \sqrt{\frac{84 + c}{c}}
\label{eq2.46}
\eeq
\beq
\alpha_{3}= \frac{2}{9 \sqrt{13}} \sqrt{\frac{18656 - 504 c - 
  11 c^2 + (272 + 11 c) \sqrt{4096 + (-1168 + c) c}}{c (7 + c)}}
\label{eq2.47}
\eeq
\beq
\alpha_{4}= \frac{2}{9 \sqrt{13}} \sqrt{\frac{18656 - 504 c - 
  11 c^2 - (272 + 11 c) \sqrt{4096 + (-1168 + c) c}}{c (7 + c)}}
\label{eq2.48}
\eeq
Substituting back, one after another, the four values of $\alpha$ in (\ref{eq2.45})-(\ref{eq2.48}) into the expression (\ref{eq2.44}) we get four allowed values of $\zeta$. Putting them back into the expressions of 15 other constants we should obtain four different solutions for the chiral algebra.

In fact, the solution with $\alpha_{2}$, in (\ref{eq2.46}), is pathological. For this value of $\alpha$, the constant $\zeta$ is zero, and $\lambda$ is also zero, which doesn't make it look like a parafermionique algebra. While some other constants are infinite. So it has to be discarded.

The solution with $\alpha_{4}$ is obtained from the solution with 
$\alpha_{3}$ by the analytic continuation in $c$. We shall explain this last equivalence slightly below.

So that, in fact, we have two independent solutions, for the constants of the parafermionic algebra. One is generated by $\alpha=\alpha_{1}$, (\ref{eq2.45}), and another one corresponding to $\alpha=\alpha_{3}$, (\ref{eq2.47}).

Resuming, we have obtained two sets of values of the constants, 
given below, for the two solutions for $\alpha$. 
We shall call them Solution 2 and Solution 3, assuming that the Solution 1 has been obtained in [2,1]. We shall denote them, symbolically as $Z_{3}^{(3.1)}$, $Z_{3}^{(3.2)}$, $Z_{3}^{(3.3)}$ parafermionic algebras.

Here, again for the notations, we mean by $Z_{3}^{(1)}$ the usual $Z_{3}$ parafermionic algebra of [3], of the 3 states Potts model, having $\Delta_{\psi}=2/3$.

By $Z_{3}^{(2)}$ we mean the parafermionic algebra of [4], of the tricritical Potts model (plus an infinite set of minimal models),
having $\Delta_{\psi}=4/3$.

Our solution for the $Z_{3}$ symmetric parafermions, we count it as the third one because it corresponds to the third value of $\Delta_{\psi}$, $\Delta_{\psi}=8/3$, allowed by the $Z_{3}$ symmetry and the associativity constraints. In fact, we have found three algebras for $\Delta_{\psi}=8/3$. So we shall denote them as $Z_{3}^{(3.1)}$, $Z_{3}^{(3.2)}$, $Z_{3}^{(3.3)}$, in the symbolic classification suggested above. In any case, as there are 
several $Z_{3}$ parafermionic algebras, they have to be called somehow, to distinguish them.

\vskip1cm

The algebras $Z_{3}^{(3.2)}$ and $Z_{3}^{(3.3)}$ correspond 
to the two sets of values of the constants which follow.

\underline{Solution 2.}

\beq
\alpha=\alpha_{1}= i \frac{8}{9} \sqrt{10} \sqrt{\frac{56 + c}{c(72 + c)}}
\label{eq2.49}
\eeq 
\beq
\zeta= \frac{4\sqrt{10}}{9} \sqrt{\frac{(272+11c)(56 + c) (112 + c)}{
  c (72 + c) (784 + 57 c)}}
\label{eq2.50}
\eeq 
\beq
\lambda= i \frac{2 \sqrt{2} (112 + c)}{9 \sqrt{c (72 + c)}}
\label{eq2.51}
\eeq 
\beq
\gamma= i \frac{2}{9} \sqrt{\frac{5}{3}} 
\sqrt{\frac{(46 + c) (56 + c) (272 + 11 c)}{
 c (72 + c) (22 + 5 c)}}
\label{eq2.52}
\eeq 
\beq
\delta= \frac{2}{27} \sqrt{70} 
\sqrt{\frac{(46 + c) (56 + c) (272 + 11 c)}{c (72 + c) (114 + 7 c)}}
\label{eq2.53}
\eeq 
\beq
\kappa= - 4 \sqrt{2} 
\sqrt{\frac{(112 + c) (272 + 11 c)}{c (72 + c) (784 + 57 c)}}
\label{eq2.54}
\eeq 
\beq
\mu= - \frac{8}{3\sqrt{3}} 
\sqrt{  \frac{(112+c)(22+5c)(46+c)}{c(72+c)(784+57c)}  }
\label{eq2.55}
\eeq 
\beq
\nu= i \frac{8}{9} \sqrt{14}(69+2c)\sqrt{ \frac{(46+c)(112+c)}{c(72+c)(114+7c)(784+57c)} }
\label{eq2.56}
\eeq 
\beq
\eta= -i \frac{\sqrt{2}}{9} \frac{(112+c)(3584+107c)}{\sqrt{c(72+c)}(784+57c)}
\label{eq2.57}
\eeq 
\beq
\tilde{\lambda}= \frac{8}{9} \sqrt{ \frac{2}{5} } 
\frac{(-33341952 + c (569296 + c (127865 + 2077 c)))\sqrt{112+c}}{\sqrt{c(56+c)(72+c)(272+11c)}(784+57c)^{3/2}}
\label{eq2.58}
\eeq 
\beq
a= i 24 \sqrt{\frac{3}{5}} 
\sqrt{ \frac{(46+c)(272+11c)}{c(72+c)(22+5c)(56+c)} }
\label{eq2.59}
\eeq 
\beq
e= -2\sqrt{\frac{21}{5}} 
\sqrt{ \frac{(72+c)(22+5c)}{c(56+c)(114+7c)} }
\label{eq2.60}
\eeq 
\beq
b= -i \sqrt{\frac{3}{5}} 
\frac{-748288 + c (10576 + c (718 + 5 c))}{\sqrt{c(46+c)(56+c)(72+c)(22+5c)(272+11c)}}
\label{eq2.61}
\eeq 
\beq
d= \frac{i}{2} \sqrt{\frac{3}{5}} 
\frac{   (2799552 + c (338980 + 7 c (1396 + 15 c))) \sqrt{(72+c)}  }{   (114+7c) \sqrt{c(22+5c)(46+c)(56+c)(272+11c)}    }
\label{eq2.62}
\eeq 
\beq
\tilde{\alpha}= -i\frac{2}{9} \sqrt{\frac{2}{5}} \frac{   2153984 + c (96736 + 1317 c)   }{   (784+57c) \sqrt{c(56+c)(72+c)}   }
\label{eq2.63}
\eeq 
\beq
\tilde{\gamma}= \frac{i}{18\sqrt{15}} 
\frac{   (738934784 + c (94531968 + c (2267608 + 15915 c))) \sqrt{46+c}   }{   (784+57c) \sqrt{c(56+c)(72+c)(22+5c)(272+11c)}   }
\label{eq2.64}
\eeq 
\beq
\tilde{\delta}= -\frac{\sqrt{7}}{27\sqrt{10}} 
\frac{   (359661568 + c (14415744 + c (67928 + 249 c))) \sqrt{46+c}   }{   (784+57c) \sqrt{c(56+c)(72+c)(114+7c)(272+11c)}   }
\label{eq2.65}
\eeq

\vskip1cm

\underline{Solution 3.}

\beq
c = 584 - 36(1/u + 65 u)\label{eq2.66a}
\eeq
With this parameterization for the central charge we have  found 
the following expressions for the constants :
\beq
\alpha=\alpha_{3}= \frac{4}{9} \sqrt{2} 
\sqrt{   \frac{u(8-45u)(6-55u)}{(1-9u)(3-20u)(9-65u)}   }
\label{eq2.66}
\eeq 
$\alpha_{3}$, as a function of $c$, is given in (\ref{eq2.47});
\beq
\zeta= 4\sqrt{\frac{2}{3}} 
\sqrt{   \frac{   (1-7u)(1-11u)(1-15u)(8-45u)(6-55u)(11-65u)   }{   (1-9u)(1-10u)(3-20u)(9-65u)(513 + u (-8518 + 33345 u))   }   }
\label{eq2.67}
\eeq 
\beq
\lambda= -\frac{2\sqrt{5}}{3} 
\frac{   (1-7u)(1-15u)   }{  \sqrt{(1-9u)(1-10u)(3-20u)(9-65u)}   }
\label{eq2.68}
\eeq 
\beq
\gamma= \frac{2}{3} 
\sqrt{   \frac{  - u(1-11u)(8-45u)(7-50u)(6-55u)(11-65u)   }{   (1-9u)(1-10u)(3-20u)(9-65u)(90+u(-1471+5850u))   }   }
\label{eq2.69}
\eeq 
\beq
\delta= \frac{2\sqrt{2}}{3\sqrt{3}} 
\sqrt{   \frac{   (2-9u)(1-11u)(7-50u)(6-55u)(11-65u)(6-65u)   }{   (1-9u)(1-10u)(3-20u)(9-65u)(126 + u (-2101 + 8190 u))   }   }
\label{eq2.70}
\eeq 
\beq
\kappa= 6\sqrt{15}
\sqrt{   \frac{   u(1-7u)(1-11u)(1-15u)(11-65u)   }{   (1-9u)(3-20u)(9-65u)(513 + u (-8518 + 33345 u))   }   }
\label{eq2.71}
\eeq 
\beq
\mu= 2\sqrt{\frac{10}{3}} 
\sqrt{   \frac{   -u(1-7u)(1-15u)(7-50u)(90 + u (-1471 + 5850 u))   }{   (1-9u)(1-10u)(3-20u)(9-65u)(513 + u (-8518 + 33345 u))   }   }
\label{eq2.72}
\eeq 
\bea
\nu= -\frac{2}{3}\sqrt{5} 
\frac{   9-104u   }{   \sqrt{(126 + u (-2101 + 8190 u))(513 + u (-8518 + 33345 u))}   }\nn\\
\sqrt{   \frac{   (1-7u)(2-9u)(1-15u)(8-45u)(7-50u)(6-65u)   }{   (1-9u)(1-10u)(3-20u)(9-65u)  }   }
\label{eq2.73}
\eea 
\beq
\eta= -\frac{\sqrt{5}}{3}
\frac{   (1-15u)(63 + u (1093 + 13 u (-2027 + 8055 u)))   }{   (513 + u (-8518 + 33345 u)) \sqrt{(1-9u)(1-10u)(3-20u)(9-65u)}  }
\label{eq2.74}
\eeq 
\bea
\tilde{\lambda}= \frac{5}{3}\sqrt{\frac{2}{3}} 
\frac{   1   }{   \sqrt{(513 + u (-8518 + 33345 u))^3}   } 
(109674 + u (-3314115 + 
    u (22473956 \nn\\
    -  5 u (-53704702 
       +  9 u (106576358 + 195 u (-2904347 + 5325840 u))))))\nn\\
\sqrt{   \frac{   (1-7u)(1-15u)   }{   (1-9u)(1-10u)(1-11u)(3-20u)(8-45u)(6-55u)(11-65u)(9-65u)   }   }          
\label{eq2.75}
\eea
\bea
a= -27\frac{(2-15u)}{\sqrt{90 + u (-1471 + 5850 u)}} \nn\\
\sqrt{   \frac{   -u(1-10u)(1-11u)(11-65u)(7-50u)   }{   (1-9u)(3-20u)(8-45u)(6-55u)(9-65u)   }   }
\label{eq2.76}
\eea 
\beq
e= 5\sqrt{3} u 
\sqrt{   \frac{   (2-9u)(6-65u)(-90 + u(1471 - 5850 u) )   }{   (1-9u)(3-20u)(6-55u)(9-65u)(126 + u (-2101 + 8190 u))   }   }
\label{eq2.77}
\eeq 
\bea
b= \frac{2232 + u (-100286 - 
     25 u (-70071 + u (598174 + 15 u (-166949 + 274950 u))))}{\sqrt{(90 + u (-1471 + 5850 u))(9-65u)(11-65u)}}\nn\\
\sqrt{   \frac{   -u   }{   (1-9u)(1-10u)(1-11u)(3-20u)(8-45u)(7-50u)(6-55u)   }   }
\label{eq2.78}
\eea 
\bea
d= \frac{5}{(252 + 2 u (-2101 + 8190 u)) \sqrt{90 + u (-1471 + 5850 u)}} 
(15876 + u (-753552 \nn\\
+   u (14595751 
     +   5 u (-29412042 
     +   13 u (12431707 + 780 u (-44861 + 49725 u)))))) \nn\\
\sqrt{   \frac{   -u(8-45u)   }{   (1-9u)(1-10u)(1-11u)(3-20u)(7-50u)(6-55u)(11-65u)(9-65u)   }   }
\label{eq2.79}
\eea
\bea
\tilde{\alpha}= -\frac{5\sqrt{2}}{9} 
\frac{(-26082 + 5 u (165057
 + u (-1921681 + 9u (1087867 - 2055105 u))))}{513 + u (-8518 + 33345 u)} 
\nn\\
\sqrt{   \frac{   u   }{   (1-9u)(3-20u)(8-45u)(6-55u)(9-65u)   }   }
\label{eq2.80}
\eea
\bea
\tilde{\gamma}= 
\frac{1}{6(513 + u (-8518 + 33345 u)) \sqrt{90 + u (-1471 + 5850 u)} } \nn\\
(987228 + 
  u (-47367636 
  +  u (922291173 + 
        5 u (-1866361932 \nn\\
        +  5 u (2071661726 
        + 585 u (-10238704 + 12079665 u)))))) \nn\\
\sqrt{   \frac{   -u(7-50u)   }{   (1-9u)(1-10u)(1-11u)(3-20u)(8-45u)(6-55u)(11-65u)(9-65u)   }   }
\label{eq2.81}
\eea
\bea
\tilde{\delta}= \frac{1}{3\sqrt{6}} 
\frac{1}{(513 + u (-8518 + 33345 u))\sqrt{126 + u (-2101 + 8190 u)}} 
(-69174 \nn\\
+  u (3206463 
  +  u (-58118544 + 
        65 u (7934262 
        + 5 u (-6909362 + 11844495 u))))) \nn\\ 
\sqrt{   \frac{   (2-9u)(7-50u)(6-65u)   }{   (1-9u)(1-10u)(1-11u)(3-20u)(6-55u)(11-65u)(9-65u)   }   }        
\label{eq2.82}
\eea 

\vskip1cm

Several remarks are in order. 

1. The Solution 2 is explicitly non-unitary. For positive values of the central charge $c$, some of the constants are imaginary, which contradicts unitarity.

We remind that in our calculations we have always assumed the two-point functions of operators to be normalized by 1, $<\psi(1)\psi^{+}(0)>=1$, etc. . So that non-unitarity shows itself not in the negative norms of the operators but in the negative values of squares of certain operator algebra constants.

Turning it differently, one can check that by the redefinition of operators :
\bea
\psi\rightarrow i\psi,\quad \psi^{+}\rightarrow -i\psi^{+},\nn\\
\tilde{\psi}\rightarrow\tilde{\psi},\quad \tilde{\psi}^{+}\rightarrow\tilde{\psi}^{+}\nn\\
U\rightarrow iU, B\rightarrow iB,  W\rightarrow W\label{eq2.83}
\eea
all operator algebra constants could be made real. But in this case the two-point functions of $U$, $B$ will become negative, positivity of the norms of the operators will be lost.

2. We assume again that all the two-point functions of the operators $\psi$, $\psi^{+}$, $\tilde{\psi}$, $\tilde{\psi}^{+}$, $U$, $B$, $W$ (forming the chiral algebra) are normalized by 1.
Still, in the definition of the operator algebra constants, in their values, there remains a liberty, a place for a choice.

It is easy to check that by changing the signs of the operators we could change the signs of several constants (cf. eq.(\ref{eq1.41})) :

$\psi\rightarrow-\psi$ and $\psi^{+}\rightarrow-\psi^{+}$ change the sign of $\lambda$ ;

$U\rightarrow-U$ change the sign of $\alpha$ ;

$B\rightarrow-B$ change the sign of $\gamma$ ;

$W\rightarrow-W$ change the sign of $\delta$ ;

$\tilde{\psi}\rightarrow-\tilde{\psi}$ and $\tilde{\psi}^{+}\rightarrow -\tilde{\psi}^{+}$ change the sign of $\zeta$ ;

In fact, in the equations of associativity, those which define the values of these five constants, they enter as squares. As a consequence there is a choice of sign for the solutions, for their values. We did made a choice, by taking a particular sign for each of them. In case of explicitly real valued expressions (for $c>0$) we have been taking them to be positive.

Once the signs of these five constants have been chosen, the rest of the constants get defined in a unique way.

3. In the case of the solution 3, equations (\ref{eq2.66})-(\ref{eq2.82}), a particular parameterization of the central charge have been used.

In the initial parameterization, by $c$, the expression for $\alpha=\alpha_{3}$, eq.(\ref{eq2.47}), contains on "internal" square root expression, which we shall denote as $h$ :
\beq
h=\sqrt{4096+(-1168c+c)c}=\sqrt{(c_{2}-c)(c_{1}-c)}\label{eq2.84}
\eeq
where
\beq
c_{1}=8(73-9\sqrt{65}),\quad c_{2}=8(73+9\sqrt{65})\label{eq2.85}
\eeq
The numbers $c_{1}$ and $c_{2}$ are both positive. 

$\alpha=\alpha_{3}$ generates the solution 3, so that the values of all the constants will contain this internal square root, eq.(\ref{eq2.84}). The expressions for the values of the constants become very long and complicated, for some of them, if the parameterization by $c$ is used.

Additional observation, which justifies the parameterization that we have used finally, was that for
\beq
0<c<c_{1}\label{eq2.86}
\eeq
($c_{1}$ is defined in (\ref{eq2.85})) all the constants are finite and real. (Some of them are divergent when $c\rightarrow 0$). The constants become complex for
\beq
c_{1}<c<c_{2}\label{eq2.87}
\eeq
and again real for
\beq
c>c_{2}\label{eq2.88}
\eeq

By analogy with the representations of the Virasoro algebra, where the first interval, containing unitary series, is
\beq
0<c<1\label{eq2.89}
\eeq
the second interval is $1<c<25$ and the third one is $c>25$, we thought that most relevant, for the present parafermionic algebra ($Z^{(3.3)}_{3}$, Solution 3) might the interval (\ref{eq2.86}).

So we have looked for the parameterization of $c$ which is natural for the interval $0<c<c_{1}$ and for which the "internal" square root 
(the expression $h$, eq.(\ref{eq2.84})) become "uniform", the internal square root would disappear.

It might be taken in the form :
\beq
c=c_{1}-36\sqrt{65}(w-2+\frac{1}{w})=584-36\sqrt{65}(w+\frac{1}{w})\label{eq2.90}
\eeq
where $w$ varies from 1 and $9/\sqrt{65}$ when $c$ varies from $c_{1}$ to $0$.
Symbolically :
\bea
w:\quad 1\rightarrow\frac{9}{\sqrt{65}},\nn\\
c :\quad c_{1}\rightarrow 0\label{eq2.91}
\eea
In this parameterization
\beq
h=36\sqrt{65}(w-\frac{1}{w})\label{eq2.92}
\eeq
and $\alpha=\alpha_{3}$, eq(\ref{eq2.47}), takes the form :
\beq
\alpha=\frac{4}{9}\sqrt{\frac{2}{13}}
\sqrt{\frac{   w(2496\sqrt{65}+w(-61256+7518\sqrt{65}w-19305w^2))   }{   7020+w(-3525\sqrt{65}+w(42802-3525\sqrt{65}w+7020w^2))   }}
\label{eq2.93}
\eeq
Otherwise, the expressions become simpler if we change $w$ for $u$ :
\beq
w=u\sqrt{65}\label{eq2.94}
\eeq
In this case
\beq
c=584-36(65u+\frac{1}{u})\label{eq2.95}
\eeq
\bea
u:\quad\frac{1}{\sqrt{65}}\rightarrow\frac{9}{65},\nn\\
c:\quad c_{1}\rightarrow 0\label{eq2.96}
\eea
\beq
h=36(65u-\frac{1}{u})\label{eq2.97}
\eeq
and $\alpha$ takes the form in (\ref{eq2.66}). We have preferred 
the parameterization by $u$, equations (\ref{eq2.66})-(\ref{eq2.82}), just because the expressions for constants become simpler.

4. Finally we observe that the Solution 4, generated by $\alpha=\alpha_{4}$ in (\ref{eq2.48}), is obtained from the Solution 3 either by the analytic continuation of $c$ (in the complex plane of $c$) around $c_{1}$, under which $h\rightarrow -h$, cf. eq.(\ref{eq2.84}), or, in the parameterization 
$w$, by the transformation
\beq
w\rightarrow\frac{1}{w}\label{eq2.98}
\eeq
or, in the parameterization $u$, by
\beq
u\rightarrow\frac{1}{65u}\label{eq2.99}
\eeq
Under the transformations (\ref{eq2.98}),(\ref{eq2.99}) the central charge is symmetric, equations (\ref{eq2.90}), (\ref{eq2.95}), while $h$ is anti symmetric, equations (\ref{eq2.92}), (\ref{eq2.97}).

\section{Classification of triple products and conclusions.}

Like in [1], after the actual calculation of the constants, in the preceding section, we shall present the general classification of the triple products, 
in an attempt to try to clarify a little bit more our method of the associativity calculations, for the present algebra.

Similarly to [1], we shall use the following classification of the chiral fields : 
$\psi,\psi^{+},\tilde{\psi},\tilde{\psi}^{+},U,B,W$.

We shall call \underline{light (l)} the fields \underline{$\psi,\psi^{+}$}; 

\underline{$U$} we shall call \underline{semi-light (l')}; 

and we shall call \underline{heavy (h)} the fields \underline{$\tilde{\psi},\tilde{\psi}^{+},B,W$}.

With this classification of the fields, the relevant, or \underline{type 1}, triple products, the products which produce equations on the constants, are those which belong to the classes :
\beq
(l,l,l),(l,l,l'),(l,l'l'),(l'l'l'),(l,l,h),(l,l',h),(l',l',h)\label{eq3.1}
\eeq

They are the products which we have already seen in the previous section.

In more detail,
\bea
\mbox{\underline{$(l,l,l)$}}:\quad \psi\psi\psi,\quad \mbox{product 3}\nn\\
\psi\psi\psi^{+},\quad \mbox{product 1}\label{eq3.2}
\eea
\bea
\mbox{\underline{$(l,l,l')$}}:\quad \psi\psi U,\quad \mbox{product 10}\nn\\
\psi\psi^{+}U,\quad \mbox{product 4}\label{eq3.3}
\eea
\beq
\mbox{\underline{$(l,l',l')$}}:\quad \psi U U,\quad \mbox{product 14}\label{eq3.4}
\eeq
\beq
\mbox{\underline{$(l',l',l')$}}:
\quad UUU,\quad \mbox{product 15}\label{eq3.5}
\eeq
\bea
\mbox{\underline{$(l,l,h)$}}:
\quad \psi\psi \tilde{\psi},\quad \mbox{product 8}\nn\\
\quad \psi\psi \tilde{\psi}^{+},\quad \mbox{product 2}\nn\\
\quad \psi\psi B,\quad \mbox{product 11}\nn\\
\quad \psi\psi W,\quad \mbox{product 12}\nn\\
\quad \psi\psi^{+}\tilde{\psi},\quad \mbox{product 9}\nn\\
\quad \psi\psi^{+}B,\quad \mbox{product 5}\nn\\
\quad \psi\psi^{+}W,\quad \mbox{product 6}\label{eq3.6}
\eea
\bea
\mbox{\underline{$(l,l',h)$}}:
\quad \psi U\tilde{\psi},\quad \mbox{product 13}\nn\\
\quad \psi U\tilde{\psi}^{+},\quad \mbox{product 7}\nn\\
\quad \psi UB,\quad \mbox{product 16}\nn\\
\quad \psi UW,\quad \mbox{product 17}\label{eq3.7}
\eea
\bea
\mbox{\underline{$(l',l',h)$}}:
\quad UU\tilde{\psi},\quad \mbox{product 19}\nn\\
\quad UUB,\quad \mbox{product 18}\nn\\
\quad UUW,\quad \mbox{product 20}\label{eq3.8}
\eea
The products in the classes
\beq
(l,h,h),(l',h,h),(h,h,h)\label{eq3.9}
\eeq
like $\psi^{+}\tilde{\psi}B$, $\psi\tilde{\psi}B$, $\psi\tilde{\psi}W$, etc., they are all irrelevant, in the sense that they do not produce equations on the operator algebra constants.

To summarize, the complete set of associativity constraints, for the present algebra, is given by the set of equations produced by products (\ref{eq3.2})-(\ref{eq3.8}). These equations are listed in the Appendix A.

To determine the operator algebra constants,  Solution 2 and Solution 3, we have used, as it has been described in the Section 2, a subset of those equations. Finally we have verified that all the remaining equations are satisfied, by the values of the constants found : (\ref{eq2.49})-(\ref{eq2.65}), Solution 2, (\ref{eq2.66})-(\ref{eq2.82}), Solution 3. This constitute the full proof of associativity of the algebras $Z_{3}^{(3.2)}$ and $Z_{3}^{(3.3)}$ presented in this paper.

As has already been stated in [1], the solutions that we have found in [2,1]
and in the present paper, the algebras $Z^{(3.1)}$, $Z^{(3.2)}$, $Z^{(3.3)}$,
they constitute the full set of solutions, of the associativity constraints.

Saying it differently, we claim that there could be no more $Z_{3}$ 
parafermionic algebras, with the dimension of the principal parafermionic fields $\Delta_{\psi}=8/3$, and having the central charge $c$ unconstrained,
remaining a free parameter.

The problem of representations of these algebras, unitary or not, 
-- the problem of physical applications, in the physicist language, 
this problem remains open.

\appendix
\section{ Equations on the operator algebra constants.}

The numeration of equations in this Appendix 
is that adopted in Section 2.
It follows the numeration of relevant triple products 
(products of the first type), considered
in Section 2 (cf. the classification in Section 3).
 
\noindent\underline{eq1.1}
\beq
  - \frac{32 (84 + c)}{81 c} + \alpha^2 + \zeta^2 - 
   \frac{8 (-116 + c) \lambda^2}{784 + 57 c} = 0
\eeq
\underline{eq1.2}
\beq
 - \frac{11 \alpha^2}{6} + \frac{4 (63168 + c (6926 + 145 c))}{
  243 c (22 + 5 c)} - \zeta^2 + \gamma^2 - \frac{(1712 + 49 c) \lambda^2}{
  784 + 57 c} = 0
\eeq
\underline{eq1.3}
\bea
 - \frac{\alpha^2 (7414 + 307 c)}{18 (114 + 7 c)} + 
  \frac{4 (333760 + c (23886 + 1025 c))}{729 c (22 + 5 c)} \nn\\
  - \delta^2 - 
  \zeta^2 + \frac{7 \gamma^2}{3} 
  - \frac{(3280 + 163 c) \lambda^2}{784 + 57 c} = 0
\eea

\noindent \underline{eq2.1}
\bea
 18 (114 + 7 c) \zeta (11 (-6272 + 1896 c + 171 c^2) \eta - 
    2 (158816 - 9052 c + 15 c^2) \lambda) \nn\\
    = (784 + 
    57 c) (4 \alpha (-93568 + 36668 c + 1851 c^2) \kappa \nn\\
    +  3 (114 + 7 c) ((-1112 + 201 c) \gamma \mu + 
       3 (-584 + 3 c) \delta \nu))
\eea
\underline{eq2.2}
\bea
6 (114 + 7 c) \zeta (13 (784 + 57 c) \eta + 
    2 (9492 + 191 c) \lambda) \nn\\
    = (784 + 
    57 c) (4 \alpha (4168 + 159 c) \kappa + 
    3 (114 + 7 c) (31 \gamma \mu + 15 \delta \nu))
\eea
\underline{eq2.3}
\bea
8 (424928 + 41086 c + 741 c^2) \kappa^2 \nn\\
+  3 (114 + 7 c) (6 (784 + 57 c) \zeta \tilde{\lambda} - 
     48 (-202 + c) \eta \lambda - (784 + 57 c) (11 \mu^2 - 6 \nu^2)) = 0
\eea

\noindent \underline{eq3.1}
\beq
(-784 - 57 c) \zeta \kappa + 9 \alpha (272 + 11 c) \lambda = 0
\eeq
\underline{eq3.2}
\beq
24 (22 + 5 c) \gamma \lambda - (784 + 57 c) \zeta \mu = 0
\eeq
\underline{eq3.3}
\beq
24 (69 + 2 c) \delta \lambda - (784 + 57 c) \zeta \nu = 0
\eeq

\noindent \underline{eq4.1}
\beq
4 a \alpha (-434 + 3 c) - 3 (114 + 7 c) (4 \delta e - 9 \alpha \gamma) = 
 0
\eeq
\underline{eq4.2}
\bea
c (-5504 (11986 + 399 c) + 
    81 \alpha^2 (22 + 5 c) (17808 + c (3793 + 132 c)) \nn\\
    -  24 a (22 + 5 c) (48 + 5 c) (784 + 57 c) \gamma + 
    12 (-42 + c) (22 + 5 c) (784 + 57 c) \kappa^2) \nn\\
    = 491925504
\eea
\underline{eq4.3}
\bea
c (405 \alpha^2 (22 + 5 c) (2256 + 53 c) - 5504 (11986 + 399 c) \nn\\
-  624 a (22 + 5 c) (784 + 57 c) \gamma + 
    544 (22 + 5 c) (784 + 57 c) \kappa^2) = 491925504
\eea
\underline{eq4.4}
\bea
16 a \alpha (137 + c) (784 + 57 c) + (114 + 
     7 c) (12 (784 + 57 c) \delta e \nn\\
     - 27 \alpha (848 + 49 c) \gamma + 
     4 (784 + 57 c) \kappa \mu) = 0
\eea
\underline{eq4.5}
\beq
135 \alpha (-48 + c) \delta + 8 (784 + 57 c) (e \gamma - \kappa \nu) = 0
\eeq

\noindent \underline{eq5.1}
\bea
2 a \alpha (784 + 57 c) (-19488 + c (-3566 + 9 c))\nn\\
=  3 (114 + 7 c) 
(4 (39 + 5 c) (784 + 57 c) \delta e \nn\\
-  9 \alpha (26544 + c (6743 + 273 c)) \gamma + 
    2 (-42 + c) (784 + 57 c) \kappa \mu)
\eea
\underline{eq5.2}
\bea
52 a \alpha (1698 + 29 c) (784 + 57 c) + (114 + 
     7 c) (396 (784 + 57 c) \delta e \nn\\
   -  27 \alpha (30224 + 1337 c) \gamma + 272 (784 + 57 c) \kappa \mu) = 0
\eea
\underline{eq5.3}
\bea
5820416 + 6 c^2 (-608 + 5174 b \gamma - 5661 \gamma^2 - 5174 \mu^2) \nn\\
+  64 c (5828 + 33 (49 b - 72 \gamma) \gamma - 1617 \mu^2) 
+  45 c^3 (38 b \gamma + 3 \gamma^2 - 38 \mu^2) = 0
\eea
\underline{eq5.4}
\bea
4 (784 + 57 c) ((114 + 7 c) d \delta - \alpha (-188 + c) e) \nn\\
= (114 + 7 c) (45 (192 + 7 c) \delta \gamma - 4 (784 + 57 c) \mu \nu)
\eea

\noindent \underline{eq7.1}
\beq
\alpha (-3056 + 147 c) \kappa - 4 (784 + 57 c) (\tilde{\alpha} \kappa + a \mu) = 0
\eeq
\underline{eq7.2}
\bea
16 a (774 + c) (784 + 57 c) \kappa + 
  3 (114 + 7 c) (4 (-1592 + 75 c) \gamma \kappa \nn\\
  -  16 (784 + 57 c) \tilde{\gamma} \kappa + (\alpha (26416 + 3 c) + 
        8 \tilde{\alpha} (784 + 57 c)) \mu + 24 (784 + 57 c) e \nu) = 0
\eea
\underline{eq7.3}
\bea
2 (4696 + 3 c) \delta \kappa - 
  4 (784 + 57 c) (2 \tilde{\delta} \kappa + e \mu) \nn\\
  + (8 \tilde{\alpha} (784 + 57 c) + \alpha (-13328 + 111 c)) \nu = 0
\eea

\noindent \underline{eq8.1}
\bea
15 \alpha (-13328 + c (1051 + 60 c)) \zeta + 
  4 (784 + 57 c) (3 \tilde{\alpha} (1 + c) \zeta \nn\\
  - (-42 + c) \eta \kappa) + 
  2 (26640 + c (-4606 + 15 c)) \kappa \lambda = 0
\eea
\underline{eq8.2}
\bea
5 \alpha (220976 + 7323 c) \zeta + 
  4 (784 + 57 c) (107 \tilde{\alpha} \zeta - 136 \eta \kappa) \nn\\
  =  32 (20681 + 324 c) \kappa \lambda
\eea
\underline{eq8.3}
\bea
7 (1024 + 27 c) \zeta \gamma + 
  2 (2 (784 + 57 c) \eta + (-2944 + 15 c) \lambda) \mu \nn\\
  =  4 (784 + 57 c) \zeta \tilde{\gamma}
\eea
\underline{eq8.4}
\bea
(9632 + 111 c) \delta \zeta + 4 (784 + 57 c) \tilde{\delta} \zeta \nn\\
=  4 (2 (784 + 57 c) \eta + 1568 \lambda - 15 c \lambda) \nu
\eea

\noindent \underline{eq9.1}
\beq
\alpha (-74 + 73 c) \kappa + 
  3 (114 + 7 c) (3 \zeta \lambda - \gamma \mu - 3 \delta \nu) = 0
\eeq
\underline{eq9.2}
\bea
3 (114 + 7 c) \zeta (10 (-224 + 3 c) (784 + 57 c) \eta + (-2257920 + 
        c (292096 + 18363 c)) \lambda) \nn\\
        + (784 + 57 c) (\alpha (1404480 + c (59438 + 6729 c)) \kappa \nn\\
        - 6 (114 + 7 c) ((-1120 + 69 c) \gamma \mu + 162 c \delta \nu)) = 0
\eea
\underline{eq9.3}
\bea
3 (114 + 7 c) \zeta (50 (784 + 57 c) \eta + 
     7 (16608 + 659 c) \lambda) \nn\\
     + (784 + 57 c) (17 \alpha (-1966 + 7 c) \kappa - 
     18 (114 + 7 c) (13 \gamma \mu + 14 \delta \nu)) = 0
\eea
\underline{eq9.4}
\bea
-361385472 + 
  c (648 \alpha \tilde{\alpha} (-961 + 2 c) (22 + 5 c) \nn\\
  + 4 (-4239776 + 75744 c - 1828332 \delta \tilde{\delta} - 1371249 \zeta^2 \nn\\
  - 1218888 \gamma \tilde{\gamma} + 3660228 \kappa^2 + 2133054 \mu^2 + 
        1828332 \nu^2) \nn\\
        +  c (-5 c (224 + 20412 \delta \tilde{\delta} + 15309 \zeta^2 + 
           13608 \gamma \tilde{\gamma} - 14580 \kappa^2 - 23814 \mu^2 - 
           20412 \nu^2) \nn\\
           - 324 (6516 \delta \tilde{\delta} + 4887 \zeta^2 + 4344 \gamma \tilde{\gamma} - 
           2 (5630 \kappa^2 + 543 (7 \mu^2 + 6 \nu^2))))) = 0
\eea
\underline{eq9.5}
\bea
278040870912 + 
  c (324 \alpha \tilde{\alpha} (22 + 5 c) (-3502 + 29 c) (784 + 57 c) + 
     1568 (358101 c \nn\\
     - 8 (-2553578 + 914166 \delta \tilde{\delta} 
     + 1828332 \zeta^2 + 
           914166 \eta^2 - 304722 \gamma \tilde{\gamma} - 761805 \kappa^2 \nn\\
           - 914166 \mu^2)) 
           +  c (-4 c (5026280 + 100176264 \delta \tilde{\delta} + 117884403 \zeta^2 + 100176264 \eta^2 \nn\\
           - 33392088 \gamma \tilde{\gamma} 
           - 83480220 \kappa^2 - 100176264 \mu^2) - 
        105 c^2 (110808 \delta \tilde{\delta} + 69741 \zeta^2 \nn\\
        + 76 (-14 + 1458 \eta^2 - 486 \gamma \tilde{\gamma} 
           - 1215 \kappa^2 - 1458 \mu^2)) - 
        972 (4263432 \delta \tilde{\delta} \nn\\
        + 7351239 \zeta^2 
        + 710572 (6 \eta^2 - 2 \gamma \tilde{\gamma} - 5 \kappa^2 - 
              6 \mu^2)))) = 0
\eea

\noindent \underline{eq13.1}
\bea
5 \alpha (1557248 + 417216 c + 4851 c^2) \zeta + 
  4 (784 + 57 c) \nn\\
  (5 \tilde{\alpha} (-224 + 3 c) \zeta - 
     4 (604 + 111 c) \eta \kappa)
     =  8 (-113344 + 3 c (45344 + 2079 c)) \kappa \lambda
\label{eqA.31}
\eea
\underline{eq13.2}
\bea
5 \alpha (314048 + 7179 c) \zeta + 
  2 (784 + 57 c) (175 \tilde{\alpha} \zeta - 487 \eta \kappa) \nn\\
  = 3 (249536 + 7853 c) \kappa \lambda
\label{eqA.32}
\eea
\underline{eq13.3}
\beq
20 \alpha (-320 + 3 c) \eta + 15 (672 + c) \zeta \kappa + 
  4 (784 + 57 c) \kappa \tilde{\lambda} = 8 \tilde{\alpha} (784 + 57 c) \eta
\label{eqA.33}
\eeq

\noindent \underline{eq18.1}
\bea
177840 + c (9552 - 84 c + 4 a^2 (-32 + c) (22 + 5 c) \nn\\
+  a b (22 + 5 c) (114 + 7 c) - (22 + 5 c) (114 + 7 c) e^2) = 0
\eea

\section{Analysis of the product $\psi\tilde{\psi}U(0)$.}

We shall analyze the expansion of the product $\psi\tilde{\psi}U(0)$ in the channel $\partial^{2}\psi^{+}$, $L_{-2}\psi^{+}$, $\tilde{\psi}^{+}$, i.e. :
\beq
\psi\tilde{\psi}U(0)\rightarrow\partial^{2}\psi^{+}(0),\quad L_{-2}\psi^{+}(0),\quad\tilde{\psi}^{+}(0)\label{eqB.1}
\eeq
The OPE of $\psi\tilde{\psi}$ is of the form :
\bea
\psi(z')\tilde{\psi}(z) = \frac{1}{(z'-z)^{\tilde{\Delta}}} 
\cdot \{\zeta\psi^{+}(z)
+(z'-z)\zeta\beta^{(1)}_{\psi\tilde{\psi},\psi^{+}} \cdot \partial\psi^{+}(z)\nn\\
+(z'-z)^{2}(\zeta\beta^{(11)}_{\psi\tilde{\psi},\psi^{+}}
\cdot\partial^{2}\psi^{+}(z)+\zeta\beta^{(2)}_{\psi\tilde{\psi},\psi^{+}}
\cdot L_{-2}\psi^{+}(z)+\eta\tilde{\psi}^{+}(z)) +...\} \label{eqB.2}
\eea

The integrals which have to be considered to derive the commutation relations $\{\psi,\tilde{\psi}\}U(0)$ are of the form (cf. [1]):
\bea
I_{1}=\frac{1}{(2\pi)^{2}}\oint_{C'_{0}}dz' (z')^{\Delta-\frac{2}{3}+n-1}\oint_{C_{0}}dz (z)^{\tilde{\Delta}-\frac{2}{3}+m-1}\nn\\
\times(z'-z)^{\tilde{\Delta}-3}\times\psi(z')\tilde{\psi}(z)U(0)\label{eqB.3}
\eea
\bea
I_{2}=\frac{1}{(2\pi)^{2}}\oint_{C_{0}}dz(z)^{\tilde{\Delta}-\frac{2}{3}+m-1}\oint_{C'_{0}}dz' (z')^{\Delta-\frac{2}{3}+n-1}\nn\\
\times(z-z')^{\tilde{\Delta}-3}\times\tilde{\psi}(z)\psi(z')U(0)\label{eqB.4}
\eea
\bea
I_{3}=\frac{1}{(2\pi)^{2}}\oint_{C_{0}}dz(z)^{\tilde{\Delta}-
\frac{2}{3}+m-1} \oint_{C_{z}}dz' (z')^{\Delta-\frac{2}{3}+n-1}\nn\\
\times\frac{1}{(z'-z)^{3}}\cdot\{\zeta\psi^{+}(z)+(z'-z)\zeta\beta^{(1)}_{\psi\tilde{\psi},\psi^{+}}\cdot\partial\psi^{+}(z)\nn\\
+(z'-z)^{2}(\zeta\beta^{(11)}_{\psi\tilde{\psi},\psi^{+}}\cdot\partial^{2}\psi^{+}(z)+\zeta\beta^{(2)}_{\psi\tilde{\psi},\psi^{+}}\cdot L_{-2}\psi^{+}(z)+\eta\tilde{\psi}^{+}(z))\}U(0)\label{eqB.5}
\eea

By the analytic continuation, one gets the relation
\beq
I_{1}+I_{2}=I_{3}
\eeq
which is transformed finally into the commutation relation 

\noindent\underline{$\{\psi,\tilde{\psi}\}U(0)$} :
\beq
\sum^{\infty}_{l=0}D^{l}_{\frac{5}{3}}\{\psi_{1+n-l}\tilde{\psi}_{-\frac{2}{3}+m+l}+\tilde{\psi}_{1+m-l}\psi_{-\frac{2}{3}+n+l}\}U(0)=R(n,m)\label{eqB.7}
\eeq
\bea
R(n,m)=\{\frac{1}{2}(\Delta-\frac{2}{3}+n-1)(\Delta-\frac{2}{3}+n-2)\zeta\psi^{+}_{-\frac{2}{3}+n+m+1}\nn\\
-(\Delta-\frac{2}{3}+n-1)\cdot(n+m+3)\cdot\zeta\beta^{(1)}_{\psi\tilde{\psi},\psi^{+}}\cdot\psi^{+}_{-\frac{2}{3}+n+m+1}\nn\\
+(n+m+4)(n+m+3)\cdot\zeta\beta^{(11)}_{\psi\tilde{\psi},\psi^{+}}\cdot\psi^{+}_{-\frac{2}{3}+n+m+1}\nn\\
+\zeta\beta^{(2)}_{\psi\tilde{\psi},\psi{+}}\cdot(L_{-2}\psi^{+})_{-\frac{2}{3}+n+m+1}+\eta\cdot\tilde{\psi}^{+}_{-\frac{2}{3}+n+m+1}\}U(0)\label{eqB.8}
\eea
$D^{l}_{\frac{5}{3}}$ in (\ref{eqB.7}) are the coefficients of the expansion :
\beq
(1-z)^{\tilde{\Delta}-3}=(1-z)^{\frac{5}{3}}=\sum^{\infty}_{l=0}D^{l}_{\frac{5}{3}}z^{l}\label{eqB.9}
\eeq
Projection of this commutation relation on the channel in which appear the operators $\partial^{2}\psi^{+}(0)$, $L_{-2}\psi^{+}(0)$, $\tilde{\psi}^{+}(0)$ requires that
\beq
-\frac{2}{3}+n+m+1=-\frac{5}{3} \quad \rightarrow \quad n+m=-2, 
\quad m=-2-n\label{eqB.10}
\eeq
With this restriction on the indexes $n,m$, the commutation relation above takes the form:
\beq
\sum^{\infty}_{l=0}D^{l}_{\frac{5}{3}}\{\psi_{1+n-l}\tilde{\psi}_{-\frac{8}{3}-n+l}+\tilde{\psi}_{-1-n-l}\psi_{-\frac{2}{3}+n+l}\}U(0)=R(n)\label{eqB.11}
\eeq
\bea
R(n)\equiv R(n,-2-n)=\{\frac{1}{2}(\Delta-\frac{2}{3}+n-1)(\Delta-\frac{2}{3}+n-2)\cdot\zeta\cdot\psi^{+}_{-\frac{5}{3}}\nn\\
-(\Delta-\frac{2}{3}+n-1)\cdot\zeta\beta^{(1)}_{\psi\tilde{\psi},\psi^{+}}\cdot\psi^{+}_{-\frac{5}{3}}\nn\\
+2\cdot\zeta\beta^{(11)}_{\psi\tilde{\psi},\psi^{+}}\cdot\psi^{+}_{-\frac{5}{3}}
+\zeta\beta^{(2)}_{\psi\tilde{\psi},\psi{+}}\cdot(L_{-2}\psi^{+})_{-\frac{5}{3}}
+\eta\cdot\tilde{\psi}^{+}_{-\frac{5}{3}}\}U(0)\label{eqB.12}
\eea
To obtain  particular commutation relations, for given values of $n$, in the explicit form, we need the explicit expressions for the actions of  modes 
of $\psi,\tilde{\psi}$ on $U(0)$. They are of the following form, as it follows from the OPEs (\ref{eq1.9}) and (\ref{eq1.12}):
\beq
\psi_{\frac{1}{3}+n}U(0)=0,\quad n>0\label{eqB.13}
\eeq
\beq
\psi_{\frac{1}{3}}U(0)=-\alpha\psi(0)\label{eqB.14}
\eeq
\beq
\psi_{-\frac{2}{3}}U(0)=-\alpha\beta^{(1)}_{\psi U,\psi}
\cdot\partial\psi(0)\label{eqB.15}
\eeq
\beq
\psi_{-\frac{5}{3}}U(0)=-\alpha\beta^{(11)}_{\psi U,\psi}\cdot\partial^{2}\psi(0)-\alpha\beta^{(2)}_{\psi U,\psi}\cdot L_{-2}\psi(0)+\kappa\tilde{\psi}(0)\label{eqB.16}
\eeq
The actions of higher modes ($\psi_{-\frac{8}{3}}U(0)$, etc.) are non-explicit. They are just descendant operators, to be dealt with, 
for their matrix elements, by using the commutation relations
of the parafermionic algebra (\ref{eqB.7}), (\ref{eqB.8}), and other commutation relations, in general, which follow from the OPEs (\ref{eq1.3})
- (\ref{eq1.20}).
\beq
\tilde{\psi}_{\frac{1}{3}+n}U(0)=0,\quad n>0\label{eqB.17}
\eeq
\beq
\tilde{\psi}_{\frac{1}{3}}U(0)=\kappa\psi(0)\label{eqB.18}
\eeq
\beq
\tilde{\psi}_{-\frac{2}{3}}U(0)=\kappa\beta^{(1)}_{\tilde{\psi}U,\psi}\cdot\partial\psi(0)\label{eqB.19}
\eeq
\beq
\tilde{\psi}_{-\frac{5}{3}}U(0)=\kappa\beta^{(11)}_{\tilde{\psi}U,\psi}\cdot\partial^{2}\psi(0)+\kappa\beta^{(2)}_{\tilde{\psi}U,\psi}\cdot L_{-2}\psi{0}-\tilde{\alpha}\tilde{\psi}(0)\label{eqB.20}
\eeq
The actions of higher modes are non-explicit.

Now, by comparing the commutation relations in (\ref{eqB.11}) and 
the explicit actions of modes in (\ref{eqB.13})-(\ref{eqB.20}) 
we find that there is just one commutation relation, 
corresponding to $n=-1$, which is going to produce an equation 
on the constants of the operator algebra. The other relations, 
for other values of $n$, will define non-explicit matrix elements, 
instead of producing equations on the operator algebra constants.

\noindent\underline{$n=-1$}:
\bea
(\psi_{0}\tilde{\psi}_{-\frac{5}{3}}+D^{1}_{\frac{5}{3}}\cdot\psi_{-1}\tilde{\psi}_{-\frac{2}{3}}+D^{2}_{\frac{5}{3}}\psi_{-2}\tilde{\psi}_{\frac{1}{3}}\nn\\
+\tilde{\psi}_{0}\psi_{-\frac{5}{3}}+D^{1}_{\frac{5}{3}}\tilde{\psi}_{-1}\psi_{-\frac{2}{3}}+D^{2}_{\frac{5}{3}}\tilde{\psi}_{-2}\psi_{\frac{1}{3}})U(0)=R(-1)\label{eqB.21}
\eea
$R(-1)$ is defined by (\ref{eqB.12}).

To transform (\ref{eqB.21}) into an equation on the constants we have to define, in an explicit form, the following matrix elements :

\noindent\underline{List 1.}
\beq
\psi^{+}_{-\frac{5}{3}}U(0),\quad(L_{-2}\psi^{+})_{-\frac{5}{3}}U(0),\quad \tilde{\psi}^{+}_{-\frac{5}{3}}U(0)\label{eqB.22}
\eeq
\underline{List 2.}
\beq
\psi_{-2}\tilde{\psi}_{\frac{1}{3}}U(0),\quad\psi_{-1}\tilde{\psi}_{-\frac{2}{3}}U(0),\quad \psi_{0}\tilde{\psi}_{-\frac{5}{3}}U(0)\label{eqB.23}
\eeq
\underline{List 3.}
\beq
\tilde{\psi}_{-2}\psi_{\frac{1}{3}}U(0),\quad\tilde{\psi}_{-1}\psi_{-\frac{2}{3}}U(0),\quad \tilde{\psi}_{0}\psi_{-\frac{5}{3}}U(0)\label{eqB.24}
\eeq
The matrix elements (\ref{eqB.22}) enter into the r.h.s. of (\ref{eqB.11}), into the expression for $R(-1)$, cf. (\ref{eqB.12}).

With some calculations one finds the expressions that follow for the matrix elements in the lists 1,2,3 above.

\noindent\underline{List 1.}
\beq
\mbox{\underline{$\psi^{+}_{-\frac{5}{3}}U(0)$}}=\alpha\beta^{(11)}_{\psi U,\psi}\cdot\partial^{2}\psi^{+}(0)+\alpha\beta^{(2)}_{\psi U,\psi}\cdot L_{-2}\psi^{+}(0)-\kappa\tilde{\psi}^{+}(0)\label{eqB.25}
\eeq
\beq
\mbox{\underline{$(L_{-2}\psi^{+})_{-\frac{5}{3}}U(0)$}}=\alpha L_{-2}\psi^{+}(0)+(\Delta+\frac{4}{3})\cdot\psi^{+}_{-\frac{5}{3}}U(0)+\alpha\beta^{(1)}_{\psi U,\psi}\cdot\partial^{2}\psi^{+}(0)\label{eqB.26}
\eeq
In the above, (\ref{eqB.26}), $\psi^{+}_{-\frac{5}{6}}U(0)$ have to be replaced by its expression in (\ref{eqB.25}). In general, everything is expressed by linear combinations of the operators $\partial^{2}\psi^{+}(0)$, $L_{-2}\psi^{+}(0)$, $\tilde{\psi}^{+}(0)$.
\beq
\mbox{\underline{$\tilde{\psi}^{+}_{-\frac{5}{3}}U(0)$}}=-\kappa\beta^{(11)}_{\tilde{\psi} U,\psi}\cdot\partial^{2}\psi^{+}(0)-\kappa\beta^{(2)}_{\tilde{\psi}U,\psi}\cdot L_{-2}\psi^{+}(0)+\tilde{\alpha}\tilde{\psi}^{+}(0)\label{eqB.27}
\eeq

\noindent\underline{List 2.}
\beq
\mbox{\underline{$\psi_{-2}\tilde{\psi}_{\frac{1}{3}}U(0)$}}=\kappa(\lambda\beta^{(11)}_{\psi\psi,\psi^{+}}\cdot\partial^{2}\psi^{+}(0)+\lambda\beta^{(2)}_{\psi\psi,\psi^{+}}\cdot L_{-2}\psi^{+}(0)+\zeta\tilde{\psi}^{+}(0))\label{eqB.28}
\eeq
\beq
\mbox{\underline{$\psi_{-1}\tilde{\psi}_{-\frac{2}{3}}U(0)$}}
=\kappa\beta^{(1)}_{\tilde{\psi}U,\psi}( (\Delta-2)\psi_{-2}\psi(0)+\lambda\beta^{(1)}_{\psi\psi,\psi^{+}}\cdot\partial^{2}\psi^{+}(0) )\label{eqB.29}
\eeq
Here,
\beq
\psi_{-2}\psi(0)=\lambda\beta^{(11)}_{\psi\psi,\psi^{+}}\cdot\partial^{2}\psi^{+}(0)+\lambda\beta^{(2)}_{\psi\psi,\psi^{+}}\cdot L_{-2}\psi^{+}(0)+\zeta\tilde{\psi}^{+}(0)\label{eqB.30}
\eeq
\beq
\mbox{\underline{$\psi_{0}\tilde{\psi}_{-\frac{5}{3}}U(0)$}}=\kappa\beta^{(11)}_{\tilde{\psi}U,\psi}\cdot\psi_{0}\partial^{2}\psi(0)+\kappa\beta^{(2)}_{\tilde{\psi}U,\psi}\cdot\psi_{0}L_{-2}\psi(0)-\tilde{\alpha}\cdot\psi_{0}\tilde{\psi}(0)\label{eqB.31}
\eeq
In the above,
\beq
\psi_{0}\partial^{2}\psi(0)=(\Delta-1)(\Delta-2)\cdot\psi_{-2}\psi(0)+2(\Delta-1)\cdot\lambda\beta^{(1)}_{\psi\psi,\psi^{+}}\cdot\partial^{2}\psi^{+}(0)+\lambda\partial^{2}\psi^{+}(0)\label{eqB.32}
\eeq
\beq
\psi_{0}L_{-2}\psi(0)=(2\Delta-2)\psi_{-2}\psi(0)+\lambda L_{-2}\psi^{+}(0)\label{eqB.33}
\eeq
\beq
\psi_{0}\tilde{\psi}(0)=\zeta\beta^{(11)}_{\psi\tilde{\psi},\psi^{+}}\cdot\partial^{2}\psi^{+}(0)+\zeta\beta^{(2)}_{\psi\tilde{\psi},\psi^{+}}\cdot L_{-2}\psi^{+}(0)+\eta\tilde{\psi}^{+}(0)\label{eqB.34}
\eeq

\noindent\underline{List 3.}
\beq
\mbox{\underline{$\tilde{\psi}_{-2}\psi_{\frac{1}{3}}U(0)$}}=-\alpha(\zeta\beta^{(11)}_{\tilde{\psi}\psi,\psi^{+}}\cdot\partial^{2}\psi^{+}(0)+\zeta\beta^{(2)}_{\tilde{\psi}\psi,\psi^{+}}\cdot L_{-2}\psi^{+}(0)+\eta\tilde{\psi}^{+}(0))\label{eqB.35}
\eeq
\beq
\mbox{\underline{$\tilde{\psi}_{-1}\psi_{-\frac{2}{3}}U(0)$}}
=-\alpha\beta^{(1)}_{\psi U,\psi}( (\tilde{\Delta}-2)\tilde{\psi}_{-2}\psi(0)+\zeta\beta^{(1)}_{\tilde{\psi}\psi,\psi^{+}}\cdot\partial^{2}\psi^{+}(0) )
\label{eqB.36}
\eeq
Here,
\beq
\tilde{\psi}_{-2}\psi(0)=\zeta\beta^{(11)}_{\tilde{\psi}\psi,\psi^{+}}\cdot\partial^{2}\psi^{+}(0)+\zeta\beta^{(2)}_{\tilde{\psi}\psi,\psi^{+}}\cdot L_{-2}\psi^{+}(0)+\eta\tilde{\psi}^{+}(0)\label{eqB.37}
\eeq
\beq
\mbox{\underline{$\tilde{\psi}_{0}\psi_{-\frac{5}{3}}U(0)$}}=-\alpha\beta^{(11)}_{\psi U,\psi}\cdot\tilde{\psi}_{0}\partial^{2}\psi(0)-\alpha\beta^{(2)}_{\psi U,\psi}\cdot\tilde{\psi}_{0}L_{-2}\psi(0)+\kappa\cdot\tilde{\psi}_{0}\tilde{\psi}(0)\label{eqB.38}
\eeq
In the above,
\beq
\tilde{\psi}_{0}\partial^{2}\psi(0)=(\tilde{\Delta}-1)(\tilde{\Delta}-2)\cdot\tilde{\psi}_{-2}\psi(0)+2(\tilde{\Delta}-1)\cdot\zeta\beta^{(1)}_{\tilde{\psi}\psi,\psi^{+}}\cdot\partial^{2}\psi^{+}(0)+\zeta\partial^{2}\psi^{+}(0)\label{eqB.39}
\eeq
\beq
\tilde{\psi}_{0}L_{-2}\psi(0)=(2\tilde{\Delta}-2)\tilde{\psi}_{-2}\psi(0)+\zeta L_{-2}\psi^{+}(0)\label{eqB.40}
\eeq
\beq
\tilde{\psi}_{0}\tilde{\psi}(0)=\eta\beta^{(11)}_{\tilde{\psi}\tilde{\psi},\psi^{+}}\cdot\partial^{2}\psi^{+}(0)+\eta\beta^{(2)}_{\tilde{\psi}\tilde{\psi},\psi^{+}}\cdot L_{-2}\psi^{+}(0)+\tilde{\lambda}\tilde{\psi}^{+}(0)\label{eqB.41}
\eeq

Finally, we have to substitute the expressions for the matrix elements, in (\ref{eqB.22}) - (\ref{eqB.41}), into the commutation relation (\ref{eqB.21}). 

The values of $\beta$-coefficients, which appear in the above expressions, are the following:
\beq
 \beta^{(1)}_{\psi\psi,\psi^{+}} = \frac{1}{2}; \quad 
 \beta^{(11)}_{\psi\psi,\psi^{+}} = \frac{128+33c}{4(784+57c)}; \quad
 \beta^{(2)}_{\psi\psi,\psi^{+}} = \frac{344}{784+57c}
\label{eqB.42}
\eeq
\beq
 \beta^{(1)}_{\psi\tilde{\psi},\psi^{+}} = \frac{1}{8}; \quad 
 \beta^{(11)}_{\psi\tilde{\psi},\psi^{+}} = \frac{5(-224+3c)}{16(784+57c)}; \quad
 \beta^{(2)}_{\psi\tilde{\psi},\psi^{+}} = \frac{350}{784+57c}
\label{eqB.43}
\eeq
\beq
 \beta^{(1)}_{\tilde{\psi}\psi,\psi^{+}} = \frac{7}{8}; \quad 
 \beta^{(11)}_{\tilde{\psi}\psi,\psi^{+}} = \frac{7(512+51c)}{16(784+57c)}; \quad
 \beta^{(2)}_{\tilde{\psi}\psi,\psi^{+}} = \frac{350}{784+57c}
\label{eqB.44}
\eeq
\beq
 \beta^{(1)}_{\tilde{\psi}\tilde{\psi},\psi^{+}} = \frac{1}{2}; \quad 
 \beta^{(11)}_{\tilde{\psi}\tilde{\psi},\psi^{+}} = \frac{11(-8+3c)}{4(784+57c)}; \quad
 \beta^{(2)}_{\tilde{\psi}\tilde{\psi},\psi^{+}} = \frac{572}{784+57c}
\label{eqB.45}
\eeq
\beq
 \beta^{(1)}_{\psi U,\psi} = \frac{7}{16}; \quad 
 \beta^{(11)}_{\psi U,\psi} = \frac{5(16+21c)}{16(784+57c)}; \quad
 \beta^{(2)}_{\psi U,\psi} = \frac{360}{784+57c}
\label{eqB.46}
\eeq
\beq
 \beta^{(1)}_{\tilde{\psi} U,\psi} = \frac{13}{16}; \quad 
 \beta^{(11)}_{\tilde{\psi} U,\psi} = \frac{346+39c)}{2(784+57c)}; \quad
 \beta^{(2)}_{\tilde{\psi} U,\psi} = \frac{402}{784+57c}
\label{eqB.47}
\eeq

Equation (\ref{eqB.21}) gives in fact 3 equations on the constants. 
This is because everything is developed in 3 independent operators, 
$\partial^{2}\psi^{+}(0)$, $L_{-2}\psi^{+}(0)$, $\tilde{\psi}^{+}(0)$. The total coefficients of these 3 operators have to vanish, to have (\ref{eqB.21}) satisfied. This gives 3 equations. With some calculations one finds, in this way, the equations 13.1, 13.2, 13.3 of the Appendix A.

One could check, using Mathematica, that all three equations 
get satisfied by the  solutions 2 and  by the solution 3, 
i.e. by the values of the constants in  (\ref{eq2.49})-(\ref{eq2.65}) 
and in (\ref{eq2.66})-(\ref{eq2.82}).

\end{document}